\documentclass[conference]{IEEEtran}

\usepackage[utf8]{inputenc}
\usepackage[T1]{fontenc}
\usepackage{graphicx}
\usepackage[hidelinks]{hyperref} 
\usepackage{url}
\usepackage{amsmath, amssymb, amsfonts}
\usepackage{booktabs} 
\usepackage[noend]{algpseudocode} 
\usepackage{algorithm}
\usepackage{listings}  
\usepackage[table]{xcolor} 
\usepackage{cite}      
\usepackage{array}     
\usepackage{longtable} 
\usepackage{rotating}  

\definecolor{codegreen}{rgb}{0,0.6,0}
\definecolor{codegray}{rgb}{0.5,0.5,0.5}
\definecolor{codepurple}{rgb}{0.58,0,0.82}
\definecolor{backcolour}{rgb}{0.98,0.98,0.98}

\lstdefinestyle{mystyle}{
    backgroundcolor=\color{backcolour},   
    commentstyle=\color{codegreen},
    keywordstyle=\color{magenta},
    numberstyle=\tiny\color{codegray},
    stringstyle=\color{codepurple},
    basicstyle=\ttfamily\footnotesize,
    breakatwhitespace=false,         
    breaklines=true,                 
    captionpos=b,                    
    keepspaces=true,                 
    numbers=left,                    
    numbersep=5pt,                  
    showspaces=false,                
    showstringspaces=false,
    showtabs=false,                  
    tabsize=2
}
\title{REMODEL-LLM: Transforming C code to Java using LLMs}

\author{
    \IEEEauthorblockN{Aryan Gupta}
    \IEEEauthorblockA{\textit{CCNSB} \\
    \textit{IIIT Hyderabad}\\
    Hyderabad, India \\
    {aryan.g@research.iiit.ac.in}}
    
    \and
    \IEEEauthorblockN{Y. Raghu Reddy}
    \IEEEauthorblockA{\textit{SERC} \\
    \textit{IIIT Hyderabad}\\
    Hyderabad, India \\
    {raghu.reddy@iiit.ac.in}}
}

\begin{document}

\maketitle
\pagestyle{plain} 

\begin{abstract}
The automated translation of C code to Java code is a notoriously difficult task, fraught with challenges stemming from fundamental paradigm shifts (procedural vs. Object Oriented), memory models (manual pointers vs.
Garbage Collection), and incompatible data types. This paper investigates the efficacy of 19 small, quantized LLMs (under 20 billion parameters) for the C to Java translation task. We use a novel, hybrid pipeline that leverages Abstract Syntax Trees (ASTs) for semantic decomposition and employs a highly constrained, rule based prompting strategy. The results are stark: a clear multi tiered performance divide emerged. The vast majority of models (Tier 3, e.g., llama3.1, gemma3, starcoder2) failed 100\% of the tests, proving incapable of generating even basic, runnable Java boilerplate. A small middle tier (Tier 2, e.g., mistral-nemo and mistral) produced runnable code but was plagued by dangerous semantic failures and wrong translations. Only three models (Tier 1: phi4, deepseek-coder-v2, codeqwen) proved viable, passing over 50\% of the test suite. Even these top models failed on the most complex C concepts, such as function pointers, sizeof, and enum logic, revealing a hard ceiling for the reasoning capabilities of current quantized models.
\end{abstract}


\section{Introduction}
Software modernization and translation remains one of the most persistent, high stakes, and costly challenges in modern software engineering. It is estimated that trillions of lines of code, a significant portion of which are written in C, still power critical infrastructure in finance, defense, aerospace, and telecommunications \cite{codex}. This C code, while often robust and highly performant, creates a significant maintenance and security burden. It is difficult to staff maintenance teams, integrate with modern cloud native architectures, and secure against contemporary vulnerabilities that prey on C's manual memory management. Translating this procedural C code to an Object-Oriented (OOP) language like Java is therefore a highly desirable, strategic goal for many organizations. A successful migration promises immense benefits, including platform independence via the Java Virtual Machine (JVM), automated memory safety through garbage collection (eliminating entire classes of buffer overflow and use after free bugs), and access to a vast, modern ecosystem of libraries, frameworks, and developers.

However, the translation from C to Java is not a simple one to one mapping of syntax, it represents a fundamental and complex paradigm shift. A successful translator must do more than swap keywords, it must semantically bridge the deep, structural chasm between the two languages. It includes the memory model, as C's manual management via {malloc()} and {free()}, along with direct pointer arithmetic, has no equivalent in Java, which uses an automated Garbage Collector (GC) and abstracted references. It also includes data types, as C's low level types like {union} (for type punning), {unsigned} integers, generic {void*} pointers, and bitfields are all designed for direct memory manipulation and do not map cleanly to Java's strict, safe type system. Furthermore, the core programming paradigm differs profoundly. C's procedural functions and global variables must be refactored into Java's class based structure, which means C {structs} must become classes, often mandating the generation of new constructors and methods. Finally, C control flow idioms, most notably the {goto} statement, are forbidden in Java and must be logically restructured into modern constructs, such as {try catch} blocks or {do while(false)} loops.

The challenge of C to Java translation is not new, and initial approaches overwhelmingly relied on rule based transpilers, also known as source to source compilers. These tools operate on a set of hand crafted rules, typically by parsing the C Abstract Syntax Tree (AST) and applying deterministic transformations to generate Java syntax. However, this approach is fundamentally flawed and brittle, especially for a translation task as complex as C to Java. This failure of deterministic, rule based systems creates the primary motivation for using Large Language Models (LLMs). LLMs, trained on billions of lines of code, operate on a probabilistic, contextual understanding \cite{lachaux2020transcoder}. Transpilers require high human maintenance cost, whereas LLMs offer zero shot flexibility. Instead of matching rigid rules, they can infer the {semantic purpose} of a code block. They offer the potential to bridge the semantic gap, handle unseen idiomatic patterns, and generate holistic, high level code structures, which is the central hypothesis of our work \cite{codex}. However, most state of the art (SOTA) code generation models are massive, proprietary, and accessible only via cloud APIs. This is often a non starter for organizations in finance, defense, or healthcare, which have strict data privacy, security, or air tight requirements and cannot risk exposing their proprietary source code to a third party API. This reality leads to a critical, unanswered question, can small, quantized, open source LLMs (e.g., Edge LLMs with fewer than 20 billion parameters) effectively handle this complex translation task? These models are fast, private, and can run entirely on local hardware. We hypothesize, however, that their compressed size and lower precision quantization (e.g., 4-bit) may render them incapable of the deep semantic reasoning required for this task. They may excel at syntactic pattern matching but fail when required to fundamentally restructure code based on a deep, contextual understanding of both C and Java.

\subsection{Related Work and Background}
The porting of legacy code is a long standing field. Early efforts focused on transpilers, or source to source compilers. These tools are almost exclusively rule based, using complex parser grammars to transform one language's syntax into another's. While effective for simple, parallel languages, they are notoriously brittle when faced with the semantic chasm between C and Java. They consistently struggle with idiomatic code, pointer logic, and the holistic architectural restructuring required to move from a procedural to an object oriented design. The limitations are not merely in implementation but are conceptual as well. Rule based tools excel at direct syntactic mapping (e.g., translating a {for} loop's syntax). They fail profoundly at bridging the {semantic chasm} between C and Java. The number of rules required to cover every C idiom, edge case, and their interactions is combinatorially explosive. Transpilers operate locally, translating one function or file at a time. They lack the holistic, architectural understanding to make intelligent refactoring decisions. For example, they cannot analyze a set of global C variables and the functions that mutate them and decide to refactor this intent into a proper Java {static} utility class or a {Singleton}. They produce a literal, non idiomatic, and often unmaintainable Java equivalent. The most complex C concepts, which are the focus of this paper, are programmatic impossibilities for most rule based systems. These include function pointers (which require a paradigm shift to Java interfaces), {union} for type punning, and complex pointer arithmetic, as these require an understanding of {programmer intent}, not just syntax.

This status quo was revolutionized by the introduction of models like OpenAI's Codex \cite{codex} and Google's AlphaCode. These models, trained on billions of lines of code, demonstrated an emergent ability to not only write code from natural language but also translate, refactor, and explain it. This success, however, is largely confined to massive, proprietary models with hundreds of billions of parameters. To address this deployment challenge, the research community has focused on two areas: model distillation for smaller, more efficient models (e.g., Mistral 7B \cite{mistral}, Llama 3.1 8B \cite{llama3_1}) and quantization \cite{quantization}. Quantization is the process of reducing the precision of a model's weights (e.g., from 32 bit floats to 4 bit integers) to dramatically reduce its memory footprint (RAM) and increase inference speed. This makes it possible to run sophisticated models on local, on premise hardware using runtimes like {ollama} \cite{ollama}. Our work specifically probes the performance degradation of a wide array of these quantized, code specific models (e.g., DeepSeek-Coder \cite{deepseek_coder_v2}, CodeLlama \cite{codellama}, StarCoder2 \cite{starcoder2}) and powerful generalist models (e.g., Gemma3 \cite{gemma3_report}, Phi-4 \cite{phi4_report}).

\section{Methodology}

Our methodology is also inspired by a growing field that combines the strengths of classical program analysis with LLMs \cite{ast_llm}. Instead of feeding a model an entire, raw source file, which is inefficient and scales poorly, new hybrid approaches use tools like Abstract Syntax Trees (ASTs) to provide a more structured, semantically rich input. By isolating a single function or {struct}, an LLM can focus its limited context window on the immediate task, and the developer can provide more targeted, guardrail driven instructions. Our pipeline, which we call REMODEL-LLM, is a direct implementation of this hybrid philosophy, using {pycparser} \cite{pycparser} for analysis and a highly constrained prompting strategy for generation.

This paper presents an empirical study to test the hypothesis that quantized LLMs lack the reasoning fidelity to bridge the C to Java gap. We achieve this by first presenting our hybrid translation pipeline, which intelligently combines classic AST based program analysis with modern LLM inference. We then detail our highly constrained, guardrail driven prompting methodology, which is designed to maximize the performance of small LLMs by explicitly stating complex translation rules. We follow this by introducing our comprehensive 20 case benchmark of C to Java translation edge cases, which are designed to probe the specific semantic weaknesses of this task. Finally, we provide a quantitative and qualitative analysis of the failure modes of 19 different quantized LLMs, demonstrating their profound limitations in semantic adherence versus syntactic correctness. This work aims to provide a realistic, data driven assessment of what is currently possible with Edge LLMs in the high stakes domain of semantic code translation.

\subsection{Models Under Test}
\begin{figure}[t] 
    \centering
    \includegraphics[width=1\linewidth]{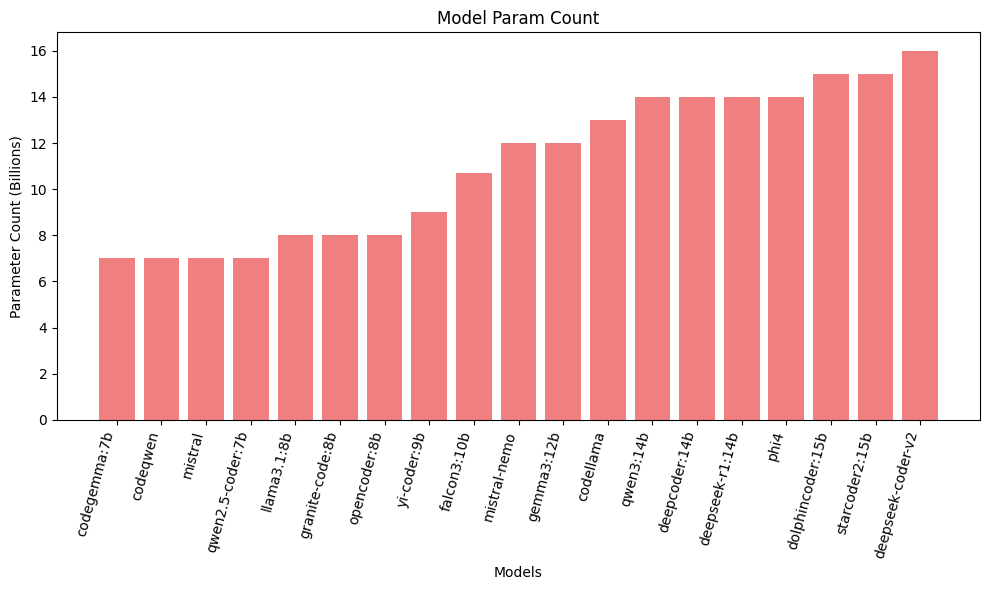}
    \caption{LLM Parameter Count}
    \label{fig:param_count}
\end{figure}

For this study, we selected a representative sample of 19 open source, decoder only LLMs with parameter counts under 20 billion. This constraint is deliberate, simulating a resource constrained Edge and on premise environment where data privacy and computational limitations prohibit the use of large, API based models. All models were run locally using the {ollama} framework \cite{ollama}. The selection was curated to include both leading code specific models and powerful generalist models to test the hypothesis that specialized training is necessary for this complex task. The parameter counts for the models tested are visualized in Figure \ref{fig:param_count}.

{Codegemma:7b} \cite{codegemma} is Google's open source model based on the Gemma 7B architecture, but specialized for code. It underwent a large, code heavy pretraining phase (500B tokens of code) followed by instruction tuning. It was included to test a strong, code native baseline from a major research lab. 

{Codeqwen} \cite{qwen1_5} is a 7B parameter variant of the Qwen1.5 family from Alibaba Cloud, specifically fine tuned for code related tasks. It is based on a Transformer architecture using SwiGLU activation and Grouped Query Attention (GQA). It was chosen to test its strong multilingual code generation and instruction following capabilities.

{qwen2.5-coder:7b} \cite{qwen1_5} is another code specific variant from Alibaba's Qwen 1.5 family. It was included to provide a direct, head to head comparison against other 7B code specific models.

{codellama:13b} \cite{codellama} is Meta's foundational 13B parameter code specific model, fine tuned from Llama 2. It was released with specialized infilling and instruction tuned variants, and serves as a critical benchmark against which all other code models are measured.

{deepseek-coder-v2} \cite{deepseek_coder_v2} is a 16B parameter sparse Mixture of Experts (MoE) model with 2.7B active parameters, making it highly efficient. It was trained on an extensive 6T token dataset of code and text, and was included as a SOTA code specific model to test the limits of MoE architectures.

{starcoder2:15b} \cite{starcoder2} is a flagship 15B parameter code model from the BigCode collaboration, led by Hugging Face and ServiceNow. It was trained on The Stack v2, a massive dataset spanning over 600 languages, and was included as a foundational, open, and permissively licensed code model.

{Dolphincoder:15b} is a popular community driven model, fine tuned from {starcoder2:15b} \cite{starcoder2} by Eric Hartford \cite{dolphincoder_hf}. It is notable for its training on a diverse, synthetic dataset for coding and its uncensored nature. It was included to represent the best of breed from open source fine tuning efforts, independent of large corporate labs.

{granite-code:8b} \cite{granite_code} is IBM's 8B parameter open source model specifically trained on code from 116 programming languages. It was designed for enterprise level code generation and modernization, making it a perfect fit for this study's objectives.

{yi-coder:9b} \cite{yi_paper} is a 9B parameter code specific model from 01.AI. It is based on the Yi architecture and was trained on over 4T tokens, with a heavy emphasis on both English and Chinese code and documentation. It was included to test the capabilities of this strong bilingual model family on a pure code logic task.

{deepcoder:14b} \cite{deepcoder_hf} represents a collaboration between the Agentica team and Together AI. It is a code reasoning model fine tuned from {Deepseek-R1-Distilled-Qwen-14B} using distributed reinforcement learning (RL). It was chosen to test the effectiveness of RL in improving complex code reasoning.

{opencoder:8b} \cite{opencoder_site} is an 8B parameter model from a community effort, notable for its training methodology. It was pretrained from scratch on 2.5 trillion tokens composed of 90\% raw code and 10\% code-related web data, and then fine tuned on 4.5M SFT examples. It was included to test this data centric, code first training approach.

{Qwen3:14b} \cite{qwen3_omni} is part of the latest generation of large language models from Alibaba's Qwen series. This family offers a comprehensive suite of dense and MoE models, and was chosen to compare the performance of a large, SOTA generalist against its smaller, code specific siblings.

{Llama3.1:8b} \cite{llama3_1} is Meta's state of the art 8B parameter generalist model. It features a GQA architecture and was pretrained on an unprecedented 15T tokens of data. It was included to serve as the SOTA benchmark for general purpose reasoning and instruction following in the <10B class.

{Gemma3:12b} \cite{gemma3_report} is Google's 12B parameter generalist model from the Gemma 3 family, built on Gemini technology. These models are multimodal, processing text and images, and feature a 128K context window. We included this model to test if its advanced multimodal architecture and long context support translate to superior reasoning on this text only, logic intensive task.

{phi4:14b} \cite{phi4_report} is Microsoft's 14B parameter state of the art model. Its uniqueness comes from its training recipe, which strategically incorporates high quality synthetic data and textbooks. It was included to test if its SOTA general reasoning, especially in STEM, would translate to this complex code logic task.

{Mistral-Nemo} \cite{mistral_nemo} is a 12B parameter small language model (SLM) representing a collaboration between Mistral AI and NVIDIA. It is optimized for high performance and efficiency. It was included to test a leading edge, commercially focused SLM's reasoning ability.

{mistral:7b} \cite{mistral} is the foundational 7B model from Mistral AI. Its architecture introduced two key innovations: Grouped Query Attention (GQA) for faster inference and Sliding Window Attention (SWA) to handle long sequences efficiently. It serves as a high performance, foundational baseline for all 7B models.

{deepseek-r1:14b} \cite{deepseek_r1} is a 14B parameter model from the DeepSeek-R1 family, which is specifically designed for reasoning. Our model variant, {DeepSeek-R1-0528}, is a distilled version fine tuned from Qwen. It was included to test the hypothesis that a model explicitly trained for general logic might outperform models trained only on code.

{falcon-3-10b} \cite{falcon_paper} is a 10.7B parameter model from TII. It was chosen to represent a different lineage of model pre training philosophies, focusing on data quality over quantity.

\subsection{Experimental Setup}

\begin{figure*}
    \centering
    \includegraphics[width=1\linewidth]{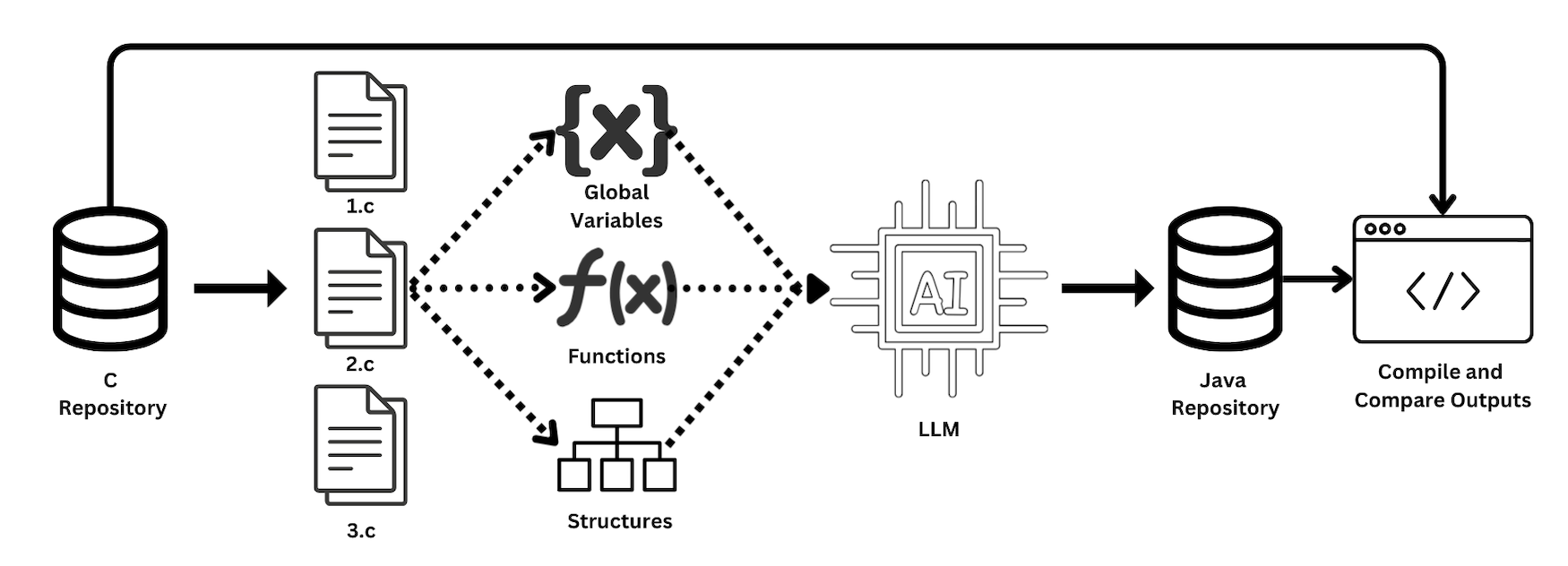}
    \caption{End to end automated C to Java translation pipeline. C source files from the repository are decomposed file wise using Abstract Syntax Tree parsing to extract global variables, functions, and data structures. These categorized code components, along with their contextual dependencies, are provided as structured prompts to a LLM for translation. The generated Java code is stored in a repository and subsequently compiled and executed to verify functional equivalence with the original C implementation through output comparison.}
    \label{fig:pipeline}
\end{figure*}

Our system, which we call REMODEL-LLM, is an end to end pipeline designed to translate and verify C code on a file by file basis (visualized in Figure \ref{fig:pipeline}). It consists of four primary stages, orchestrated by a central {PipelineManager}: C Code Analysis, Prompt Driven Translation, Post Processing and Assembly, and Automated Verification. This hybrid approach is designed to supplement the generative capabilities of LLMs with the precision of classical program analysis.

The pipeline begins with the {CAnalyzer} module. This module is built on the {pycparser} library \cite{pycparser} and, instead of naive text parsing, constructs a full Abstract Syntax Tree (AST) of the C code. We then traverse this AST using a custom {ContextVisitor}. This AST based approach is critical because it allows us to semantically decompose the C file into its constituent parts: global variables, struct and union definitions, function prototypes, and function implementations. Each node in the AST contains its exact source code. The output of this stage is a structured dictionary containing lists of code snippets for each category, which are then passed to the translator.

The core of our system is the {JavaTranslator} module. This module is responsible for orchestrating calls to the LLMs via the {ollama} library \cite{ollama}. It does not translate the entire file at once. Instead, it iterates through the decomposed parts from the analysis stage and uses one of three specialized, guardrail driven prompts. Our key insight is that small, quantized models cannot be trusted to infer complex translation rules and must be explicitly instructed. The full text of these prompts is available in Listings \ref{lst:func_prompt}, \ref{lst:global_prompt}, and \ref{lst:struct_prompt}. This guardrail driven methodology is one of the main contribution of our approach. For example, the function prompt (Listing \ref{lst:func_prompt}) explicitly forbids the {goto} keyword, instructing the model to restructure C's goto cleanup idiom into a Java {do while(false)} and {break} pattern. It also forbids the {union} keyword, providing a rule to translate its usage directly using Java's bit converters like {Float.floatToIntBits}. Similarly, it forbids the {unsigned} keyword, mandating a translation to {long} with {\& 0xFFFFFFFFL} masking to simulate 32 bit wrapping. It also provides rules for handling pointer arithmetic, struct pass by value semantics, and macro inlining. Separate, simpler prompts (Listing \ref{lst:global_prompt} and \ref{lst:struct_prompt}) are used for translating global variables into {public static} fields and {structs} into {public static} nested classes with mandatory constructors.

Once the LLM generates a translation, it passes to the {Post Processing} stage. LLMs, especially chat models, are prone to hallucinating conversational text, markdown formatting, and extra code. Our {JavaTranslator} implements a robust post processing chain to clean the raw output. This includes a regex based function to strip java blocks and conversational intros. A second function, {\_strip\_imports}, removes any hallucinated {import} statements. Finally, a critical function, {\_strip\_extra\_definitions}, performs a brace count search to find and extract only the single function we asked for, discarding any extra hallucinated methods or classes. The {PipelineManager} then assembles these cleaned Java snippets, globals, nested classes, and methods, into a single, syntactically valid Java file.

The final stage is {Automated Verification}, handled by the {CodeExecutor}. This module provides the binary pass/fail ground truth. It first compiles and runs the original C file using gcc, capturing its {stdout}. It then compiles and runs the newly generated Java file using javac and java, capturing its {stdout}. Finally, it performs a strict string equality comparison on the outputs from both processes. If the outputs match exactly, the translation is marked as a success. If they differ, or if either file fails to compile or run, it is marked as a failure.

\begin{lstlisting}[language=Python, basicstyle=\ttfamily\scriptsize, breaklines=true, caption={The primary guardrail prompt for C function translation, detailing explicit rules for the semantic conversion of complex idioms (e.g., {goto}, {union}, pointers, and {unsigned} types).}, label={lst:func_prompt}]
self.function_prompt_template = You are a C-to-Java translation engine. Your task is to translate *only* the single C function provided.

**TASK:**
Translate the C function below into a single, syntactically correct Java method.

**C Function:**
{function_code}

**CONTEXT (For Type Information Only):**
- C Globals (if any): {globals_context}
- C Structs (if any): {structs_context}

**CRITICAL RULES. YOU MUST FOLLOW THESE:**
1.  **SINGLE FUNCTION ONLY:** Your *entire* response MUST be *only* the Java method.
    * **DO NOT** write import statements.
    * **DO NOT** write class definitions.
    * **DO NOT** include other methods, even if this function calls them.
    * **DO NOT** write explanations or markdown.
2.  **ILLEGAL KEYWORDS & PATTERNS (DO NOT USE):**
    * **goto (CRITICAL):** DO NOT write the keyword goto. C code using goto for cleanup (e.g., if (err) goto cleanup; ... cleanup: ...) MUST be restructured:
        1.  You MUST wrap the *relevant* C code in a Java do_while_loop: do {{ ... }} while(false); block.
        2.  The C goto cleanup; line MUST be translated to break do_while_loop;.
        3.  The C cleanup: label itself is translated as the code *after* the }} while(false); block.
    * **union (CRITICAL):** DO NOT write the keyword union.
        1.  Any C line that *declares* a union variable (e.g., union Converter c;) MUST be **DELETED**.
        2.  When that variable is *used* (e.g., c.f = 1.2f; int i = c.i;), you MUST translate this usage directly using Java's bit converters (e.g., float f_val = 1.2f; int i = Float.floatToIntBits(f_val);).
    * **unsigned (CRITICAL):** DO NOT write the keyword unsigned. Translate unsigned int to long and mask all arithmetic with & 0xFFFFFFFFL.
    * **& (Address-Of Operator - CRITICAL):** DO NOT use the & operator.
        1.  When passing a variable by pointer (e.g., move(&p1);), you MUST translate this to pass the Java reference directly: move(p1);.
        2.  (The *only* exception is C's scanf, which is not in these tests).
3.  **main FUNCTION:** If the C function name is 'main', the Java signature MUST be public static void main(String[] args). All other functions MUST be public static.
4.  **printf (CRITICAL):** String literals inside printf (e.g., Overflow ( + 1):) MUST be preserved **VERBATIM**.
    * Copy the string *exactly* as it appears in C.
    * **DO NOT** fix or correct spacing, typos, or formatting.
5.  **POINTER ARITHMETIC:**
    * int *p = arr; -> int p_index = 0; int[] p_array = arr;
    * *(p + 2) -> p_array[p_index + 2].
    * p++ -> p_index++.
6.  **STRUCT PASS-BY-VALUE:**
    * A C function void move_copy(Point p_val) receives a *copy*.
    * The Java method MUST create a copy: public static void move_copy(Point p_val_orig) {{ Point p_val = new Point(p_val_orig); p_val.x += 100; }}. (This assumes a copy constructor Point(Point other) exists).
7.  **MACRO INLINING:**
    * Macros with side effects MUST be inlined.
    * int k = MAX(i++, j++); -> int k = ((i++) > (j++) ? (i++) : (j++));

\end{lstlisting}

\begin{lstlisting}[language=Python, basicstyle=\ttfamily\scriptsize, breaklines=true, caption={The specialized prompt for translating C global variables into {public static} Java class fields, including type mapping rules for {const} and {unsigned}.}, label={lst:global_prompt}]
self.global_prompt_template = You are a C to Java translation expert. Your task is to convert the following C global variable into a 'public static' Java class field. Maintain the type and initial value.

**CRITICAL TRANSLATION RULES:**
1.  **Type:** Map C types to Java: int -> int, float -> float, char* -> String.
2.  **const:** const variables MUST be translated to final in Java.
3.  **unsigned:** unsigned int MUST be translated to long. unsigned char MUST be translated to int.
4.  **Visibility:** All globals become 'public static'.
5.  **Output:** Provide ONLY the single line of Java code.

The C code to translate is:
{c_code}

\end{lstlisting}

\begin{lstlisting}[language=Python, basicstyle=\ttfamily\scriptsize, breaklines=true, caption={The prompt for mapping C global variables to {public static} Java fields, enforcing type and modifier conversion.}, label={lst:struct_prompt}]
self.struct_prompt_template = You are a C to Java translation expert. Your task is to convert the following C struct or union definition into a 'public static' nested Java class.

**CRITICAL TRANSLATION RULES:**
1.  **union (CRITICAL):** If the C code provided is a union, you MUST output **NOTHING**. No text, no code, no comments. Just an empty string.
2.  **struct (Standard):** If the code is a struct, translate it to a public static Java class.
    * All C fields should become public Java fields.
    * char name[50]; -> public char[] name = new char[50];
    * char *name; -> public String name;
3.  **CONSTRUCTORS (MANDATORY):** Your Java class MUST include all three of these constructors:
    1.  A public no-argument default constructor.
    2.  A public copy constructor that takes one argument (e.g., public Point(Point other)).
    3.  A public all-field constructor that takes arguments for *all* fields defined in the struct.
        * (e.g., for Point {{ int x; int y; }}, you MUST add public Point(int x, int y) {{ this.x = x; this.y = y; }}).
4.  **Bitfields:** struct Flags {{ int a:1; }} MUST be translated to a class with private data and public getters/setters using bit-masking.
5.  **Output:** Provide ONLY the Java class definition. Do not include import statements or any other text.

The C code to translate is:
{c_code}

\end{lstlisting}

\begin{table*}[t!]
    \centering
    \caption{C-to-Java Idiomatic Test Suite}
    \label{tab:test_suite}
    \scriptsize
    \renewcommand{\arraystretch}{1.1} 
    \begin{tabular}{@{}ll@{}}
        \toprule
        \textbf{Test Case} & \textbf{Description / C Idiom Tested} \\
        \midrule
        Test 1 & Tests translation of basic pointer arithmetic (p++, \*(p+2)) into Java array index manipulation (p\_index++, arr[p\_index + 2]). \\
        Test 2 & Tests C's struct pass by value (copying) vs. pass by pointer (reference), requiring Java to simulate copying with a new Point(other) constructor. \\
        Test 3 & Tests C union for type punning (float/int bits), requiring translation to Java's bit level converters (e.g., Float.floatToIntBits). \\
        Test 4 & Tests C preprocessor macro inlining, specifically the dangerous MAX(i++, j++) side effect which must be preserved literally. \\
        Test 5 & Tests unsigned int arithmetic, checking for correct translation to Java long with \& 0xFFFFFFFFL masking to simulate 32 bit wrapping. \\
        Test 6 & Tests goto for forward jumping error cleanup, requiring a structural refactor into a do while(false) and break pattern. \\
        Test 7 & Tests C function pointers, requiring a major paradigm shift to Java's functional interfaces (e.g., IntBinaryOperator). \\
        Test 8 & Tests struct bitfields (int a:1), requiring translation to a Java class with private fields and manual bit masking getters/setters. \\
        Test 9 & Tests C's generic void\* pointers, requiring translation to Java's Object type with explicit runtime casting. \\
        Test 10 & Tests C style out parameters using a pointer to a pointer (int\*\*), requiring a refactor to a Java method that returns the value. \\
        Test 11 & Tests unsigned char (0-255 range), which must be mapped to a Java int and masked (\& 0xFF) to avoid sign extension issues. \\
        Test 12 & Tests C enum treated as an integer (s = s + 1), a semantic trap that is illegal in Java and requires complex refactoring. \\
        Test 13 & Tests goto for a backward jumping loop, requiring a structural refactor into a Java while or do while loop. \\
        Test 14 & Tests C string.h functions (strcpy, strcat), requiring translation from char[] manipulation to Java String object operations. \\
        Test 15 & Tests C dynamic memory allocation (malloc, free), requiring translation to Java's new operator and reliance on the Garbage Collector. \\
        Test 16 & Tests mutable global variables, which must be translated to public static Java fields. \\
        Test 17 & Tests a simple, multi line C macro, checking for correct inlining of the code block. \\
        Test 18 & Tests the C specific sizeof operator, which has no Java equivalent and must be refactored or removed by the LLM. \\
        Test 19 & Tests nested struct access (outer.inner.value) and initialization, which must be mapped to nested Java classes. \\
        Test 20 & Tests a switch statement with C style fall through, a control flow that maps directly to Java's switch (if break is omitted). \\
        \bottomrule
    \end{tabular}
\end{table*}

To evaluate the semantic understanding of the models, we designed a challenging test suite of 20 C files. This benchmark is not designed to test bulk translation, but to probe specific, difficult to translate C idioms that represent the core C to Java chasm. The test cases are categorized in Table \ref{tab:test_suite}. The primary metric for our evaluation is the binary verification pass rate provided by the {CodeExecutor} module. For each model, we report the percentage of the 20 test cases that successfully compiled, ran, and produced a {stdout} identical to the C baseline. A pass (P) indicates a perfect match. A fail (F) indicates any error, including a failure to compile the generated Java code, a runtime exception (e.g., {NullPointerException}), or a semantic failure where the Java code ran but produced an output that did not match the C baseline. This strict, correctness based metric is essential for a high stakes task like code translation.

\section{Results}
\label{sec:results}

The results of our comprehensive evaluation are summarized in Table \ref{tab:results_summary}. Each of the 19 models was run against the 20 case benchmark, and the performance data reveals a clear all or nothing divide, where the majority of models failed completely, and only a select few demonstrated partial success, forming distinct tiers of competence.

\begin{table*}[t!]
    \centering
    \caption{Full C-to-Java Translation Results (P=Pass, F=Fail)}
    \label{tab:results_summary}
    \scriptsize
    \setlength{\tabcolsep}{5pt} 
    \begin{tabular}{@{}l|c|cccccccccccccccccccc@{}}
        \toprule
        {Model} & {Total} & {T1} & {T2} & {T3} & {T4} & {T5} & {T6} & {T7} & {T8} & {T9} & {T10} & {T11} & {T12} & {T13} & {T14} & {T15} & {T16} & {T17} & {T18} & {T19} & {T20} \\
        \midrule
        codegemma:7b & {0} & F & F & F & F & F & F & F & F & F & F & F & F & F & F & F & F & F & F & F & F \\
        codeqwen & {11} & P & P & F & P & P & P & F & F & P & F & P & F & P & F & P & F & P & F & F & P \\
        mistral-nemo & {8} & P & P & F & P & F & F & F & F & F & F & P & F & F & P & P & F & P & F & F & P \\
        llama3.1:8b & {0} & F & F & F & F & F & F & F & F & F & F & F & F & F & F & F & F & F & F & F & F \\
        gemma3:12b & {0} & F & F & F & F & F & F & F & F & F & F & F & F & F & F & F & F & F & F & F & F \\
        dolphincoder:15b & {0} & F & F & F & F & F & F & F & F & F & F & F & F & F & F & F & F & F & F & F & F \\
        mistral & {6} & P & F & F & F & F & F & F & F & P & F & F & F & P & P & F & F & P & F & F & P \\
        qwen2.5-coder:7b & {0} & F & F & F & F & F & F & F & F & F & F & F & F & F & F & F & F & F & F & F & F \\
        falcon3:10b & {0} & F & F & F & F & F & F & F & F & F & F & F & F & F & F & F & F & F & F & F & F \\
        granite-code:8b & {0} & F & F & F & F & F & F & F & F & F & F & F & F & F & F & F & F & F & F & F & F \\
        yi-coder:9b & {0} & F & F & F & F & F & F & F & F & F & F & F & F & F & F & F & F & F & F & F & F \\
        deepcoder:14b & {0} & F & F & F & F & F & F & F & F & F & F & F & F & F & F & F & F & F & F & F & F \\
        opencoder:8b & {0} & F & F & F & F & F & F & F & F & F & F & F & F & F & F & F & F & F & F & F & F \\
        deepseek-r1:14b & {0} & F & F & F & F & F & F & F & F & F & F & F & F & F & F & F & F & F & F & F & F \\
        deepseek-coder-v2 & {13} & P & P & P & P & P & P & F & P & F & F & P & F & P & P & F & P & P & F & F & P \\
        qwen3:14b & {0} & F & F & F & F & F & F & F & F & F & F & F & F & F & F & F & F & F & F & F & F \\
        starcoder2:15b & {0} & F & F & F & F & F & F & F & F & F & F & F & F & F & F & F & F & F & F & F & F \\
        codellama & {0} & F & F & F & F & F & F & F & F & F & F & F & F & F & F & F & F & F & F & F & F \\
        phi4 & {11} & P & P & F & P & P & F & F & F & P & F & P & F & P & P & F & F & P & P & F & P \\
        \bottomrule
    \end{tabular}
    \flushleft
\end{table*}

\subsection{Per Test Case Analysis}

\begin{figure}[t] 
    \centering
    \includegraphics[width=1\linewidth]{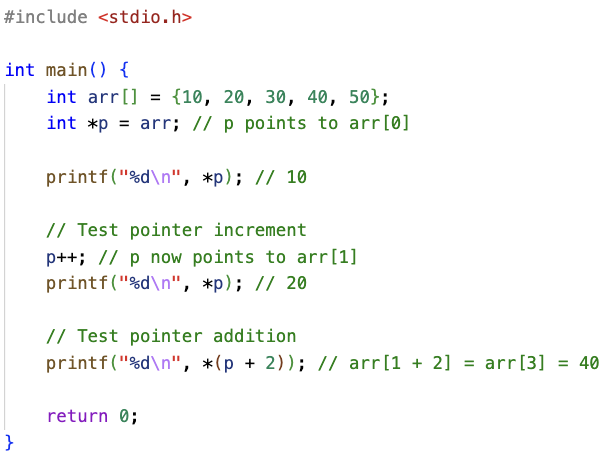}
    \caption{Test Case 1 (test\_1.c), a foundational and easy test for C pointer arithmetic. This test checks the model's ability to translate fundamental pointer operations. Initialization (int \*p = arr;), incrementing (p++), and offset dereferencing (\*(p + 2)). The ideal solution must refactor this C specific logic into Java array index manipulation, using a separate index variable (e.g., int p\_index = 0;) and replacing all pointer operations with array lookups (e.g., p\_array[p\_index] and p\_array[p\_index + 2]).}
    \label{fig:tc1}
\end{figure}

\begin{figure}[t] 
    \centering
    \includegraphics[width=1\linewidth]{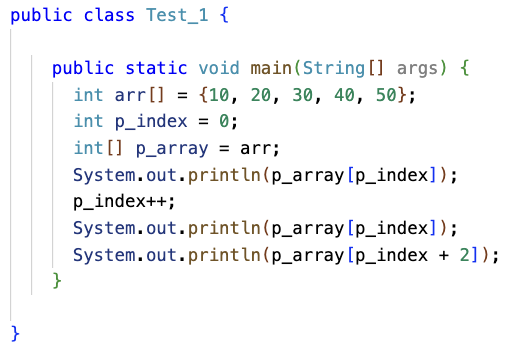}
    \caption{The successful translation of test\_1.c by deepseek-coder-v2. This output demonstrates a perfect adherence to the POINTER ARITHMETIC rule provided in the prompt. The model correctly refactored the C pointer (int \*p) into a separate Java index variable (int p\_index = 0) and an array reference (int[] p\_array). Consequently, it translated C's \*p, p++, and \*(p + 2) into the semantically equivalent Java operations p\_array[p\_index], p\_index++, and p\_array[p\_index + 2], leading to a passing test.}
    \label{fig:tc1_out}
\end{figure}

Analyzing the pass rate for each specific test case, as shown in Table \ref{tab:test_case_analysis}, provides deep insight into the types of reasoning that quantized models struggle with. We can group the tests into categories of difficulty, which map directly to the semantic gap between C and Java. For example, Test Case 1 (T1) was a foundational test for pointer arithmetic (shown in Figure \ref{fig:tc1}). The successful output from {deepseek-coder-v2} (Figure \ref{fig:tc1_out}) demonstrates a perfect adherence to the prompt's rules for this task, refactoring pointer logic into index logic. In sharp contrast, Test Case 13 (T13) presented a much harder C specific control flow problem, requiring the model to refactor a goto based loop (Figure \ref{fig:tc13_src}). This task prompted a characteristic failure from a Tier 2 model, as shown in Figure \ref{fig:tc13_fail}. The model, mistral-nemo, produced a syntactically invalid translation by naively mapping goto to continue, demonstrating its inability to perform the required semantic restructuring. This juxtaposition of a simple success and a complex failure highlights the clear reasoning ceiling of most models.

\begin{figure}
    \centering
    \includegraphics[width=1\linewidth]{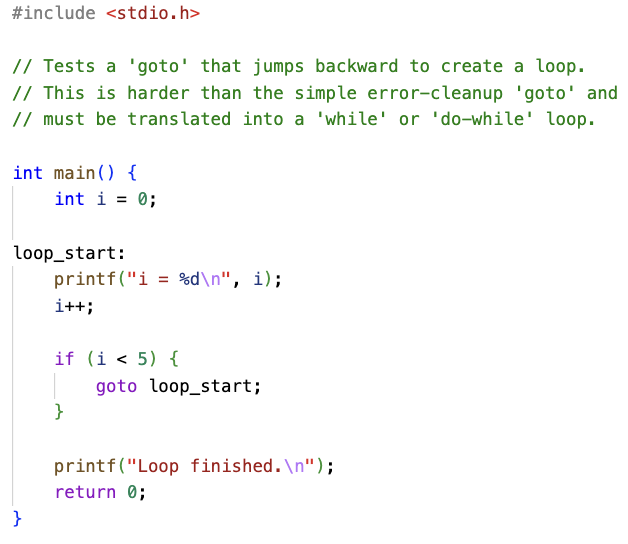}
    \caption{Test Case 13 (test\_13.c), a C specific control flow test. This case probes a model's ability to refactor a backward jumping goto statement, which creates a loop. A semantically correct translation requires the model to understand the C code's intent and restructure this pattern into a standard Java while or do-while loop, as Java does not support goto for control flow.}
    \label{fig:tc13}
\end{figure}

\begin{figure}
    \centering
    \includegraphics[width=1\linewidth]{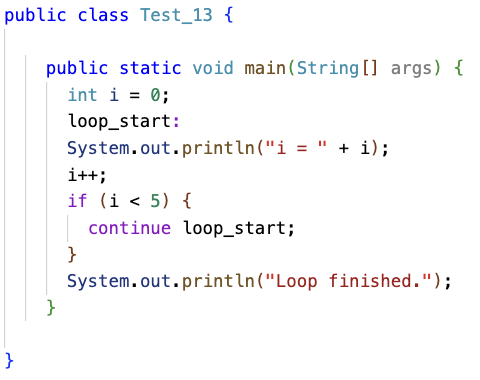}
    \caption{A characteristic failure on Test 13 by mistral-nemo. This output demonstrates a literal syntax translation failure. The model has incorrectly mapped C's goto loop\_start to Java's continue loop\_start. This is syntactically invalid in Java, as a labeled continue statement is only permitted inside an enclosing loop structure, which is absent here. The code fails to compile, highlighting the model's shallow, pattern matching based reasoning and its inability to perform the necessary semantic restructuring.}
    \label{fig:tc13_out}
\end{figure}

\begin{table*}[t!]
    \centering
    \caption{Per-Test Case Analysis (19 Models)}
    \label{tab:test_case_analysis}
    \renewcommand{\arraystretch}{1.1}
    \begin{tabular}{@{}lccl@{}}
        \toprule
        {Test} & {Pass} & {Pass} & {Analysis \& Insight} \\
        {Case} & {Count} & {Rate} & \\
        \midrule
        \multicolumn{4}{l}{\textit{{The Impossible Tests: 0\% Success}}} \\
        T7 & 0 & 0\% & {Function Pointers:} Requires full semantic change to Java interfaces. All models failed. \\
        T10 & 0 & 0\% & {Pointer to Pointer ({int\*\*}):} Requires refactoring to a return value or wrapper. All models failed. \\
        T12 & 0 & 0\% & {Enum as Int:} A semantic trap. {s = s + 1;} is invalid in Java and requires complex logic. All models failed. \\
        T18 & 0 & 0\% & {{sizeof} Operator:} No Java equivalent. No model could reason a valid removal/replacement. \\
        T19 & 0 & 0\% & {Nested Structs:} Simple task, but nested object access and C style string handling (strcpy) overwhelmed all models. \\
        \midrule
        \multicolumn{4}{l}{\textit{{The Nearly Impossible Tests: 1 Model Passed}}} \\
        T3 & 1 & 5\% & {Union:} Passed only by {deepseek-coder-v2}. Prompt rule to use was {Float.floatToIntBits}. \\
        T8 & 1 & 5\% & {Bitfields:} Passed only by {deepseek-coder-v2}.  Prompt rule to use was to create bit masked getters/setters. \\
        \midrule
        \multicolumn{4}{l}{\textit{{The Hard Tests: Low Success}}} \\
        T6 & 2 & 11\% & {{goto} (cleanup):} Only {deepseek} and {codeqwen} followed the {do-while(false)/break} rule. \\
        T15 & 2 & 11\% & {{malloc}/{free}:} Most models failed to replace with {new}. \\
        T16 & 2 & 11\% & {Globals:} Most models failed to map to {static}. \\
        T5 & 3 & 16\% & {{unsigned int}:} Only 3 models followed the {long} and {\& 0xFFFFFFFFL} mask rule. \\
        \midrule
        \multicolumn{4}{l}{\textit{{The Moderate Tests: High Variance}}} \\
        T2 & 4 & 21\% & {Struct Pass by Value:} Required creating a copy constructor. \\
        T9 & 4 & 21\% & {{void\*}:} Required translation to {Object} and casting. \\
        T14 & 4 & 21\% & {{string.h}:} Required mapping {strcpy}/{strcat} to Java {String} operations. \\
        T4 & 5 & 26\% & {Macro (side effect):} Models had to inline {((i++) > (j++) ? (i++) : (j++))}. \\
        T11 & 5 & 26\% & {{unsigned char}:} Required mapping to {int} and {\& 0xFF}. \\
        T13 & 5 & 26\% & {{goto} (loop):} Models were able to refactor this into a {while} loop. \\
        T17 & 5 & 26\% & {Multi line Macro:} Simple text expansion. \\
        T1 & 5 & 26\% & {Pointer Arithmetic:} Required mapping to array indices. \\
        T20 & 5 & 26\% & {{switch} fall through:} Direct 1:1 syntax mapping. \\
        \bottomrule
    \end{tabular}
\end{table*}

The first category, The Impossible Tests, consists of five test cases with a 0\% success rate across all 19 models, as detailed in Table \ref{tab:test_case_analysis}. These failures highlight the hard ceiling of current models. They include {T7 (Function Pointers)} and {T10 (Pointer to Pointer)}, both of which require a major paradigm shift from C's pointer based semantics to Java's object oriented (interface based) or value returning patterns. Models also universally failed {T18 (sizeof Operator)}, which has no Java equivalent and requires abstract reasoning to remove or replace. {T12 (Enum as Int)} proved to be a semantic trap where the C idiom {s = s + 1;} is syntactically illegal in Java, and no model could refactor the logic. Finally, {T19 (Nested Structs)} was a surprising 0\% failure, which likely failed due to the combined complexity of nested object access and C style string manipulation ({strcpy}).

The second category, The Nearly Impossible Tests, was passed by only one model. As seen in Table \ref{tab:test_case_analysis}, both {T3 (Union)} and {T8 (Bitfields)} were passed only by {deepseek-coder-v2}. This is a critical finding. It proves that this model was uniquely capable of reading, understanding, and correctly applying the complex, specific rules provided in our prompts (e.g., using {Float.floatToIntBits} for the union and creating bit masked getters/setters for the bitfield). This suggests a superior in context learning and reasoning capability.

A set of Hard tests revealed low success rates across the board. These include {T6 (goto cleanup)}, passed only by the two models that correctly followed the complex {do-while/break} refactoring rule from the prompt. {T15 (malloc/free)} and {T16 (Globals)} were also passed by only two models, showing a general failure to map C's memory and state management to Java's {new} operator and {static} fields. {T5 (unsigned int)} was passed by only three models, as most failed to apply the {long} and bit masking rule.

The remaining Moderate tests, also detailed in Table \ref{tab:test_case_analysis}, showed high variance and served as a baseline for basic competence. Simple pointer arithmetic ({T1}), switch fall through ({T20}) and others were the easiest, passed by 5 of 19 models, as their logic maps relatively directly to Java array indices and {switch} blocks. Tasks like macro inlining ({T2}) and string functions ({T14}) were only handled by the best performing models.

\subsection{Overall Model Performance Tiers}
\begin{figure}[t] 
    \centering
    \includegraphics[width=1\linewidth]{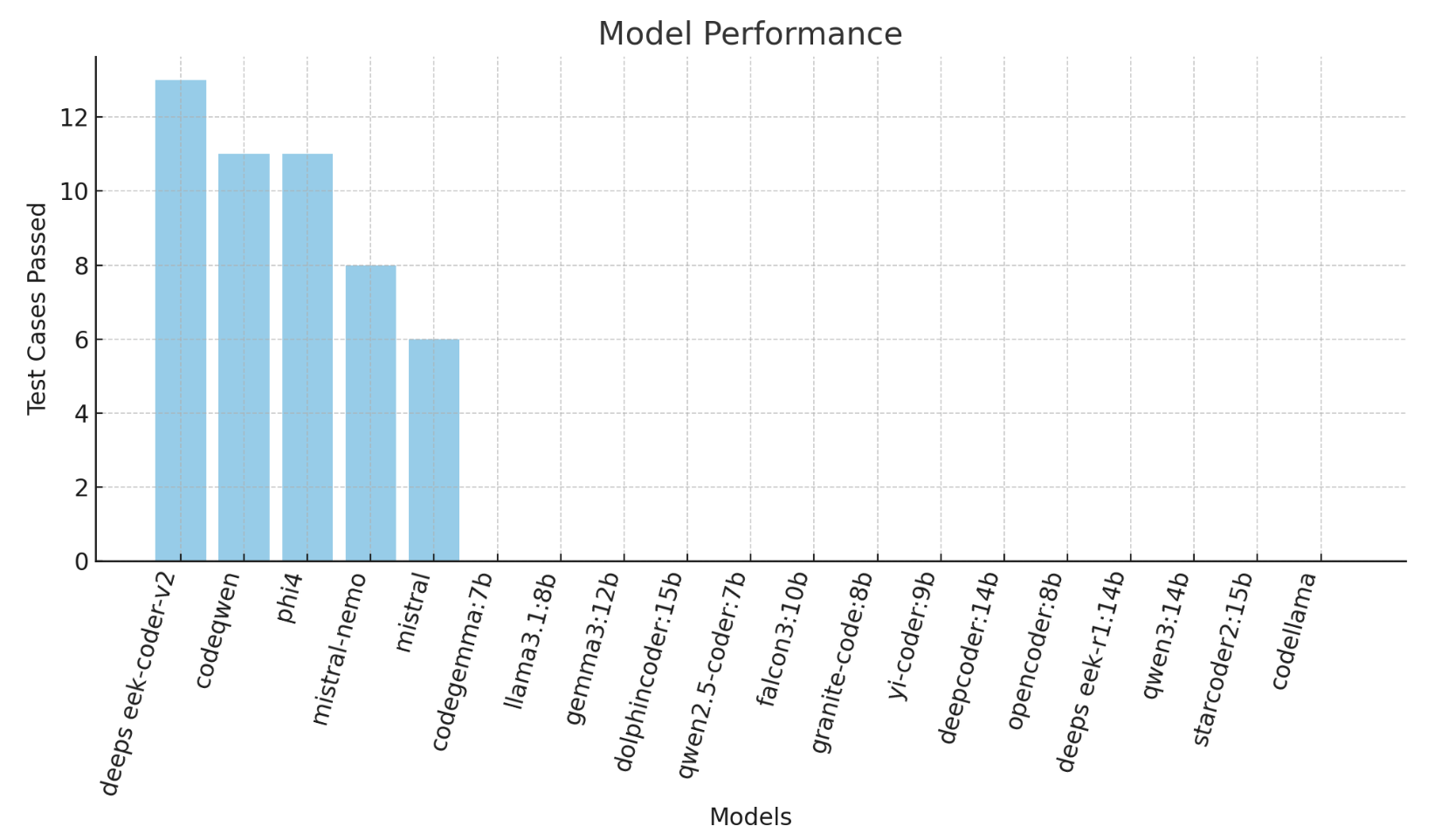}
    \caption{Total number of test cases passed (out of 20) by each of the 19 LLMs. The results highlight a stark, multi tiered performance divide on the C to Java translation task.}
    \label{fig:Model_Perf}
\end{figure}

A stark, multi tiered difference in model performance is immediately evident, as illustrated in Figure \ref{fig:Model_Perf}. We categorize the models into three distinct tiers based on their ability to handle the translation task.

The first, Tier 1 (Viable), contains only three models: {{deepseek-coder-v2}} (13/20), {{codeqwen}} (11/20) and {{phi4}} (11/20). These models are in a class of their own, successfully compiling and passing over 50\% of the tests, as shown in the totals column of Table \ref{tab:results_summary}. Their success indicates a clear, though incomplete, ability to understand and act on the complex, rule based instructions provided in our prompts. Their failures were not on basic syntax, but on the most advanced C specific concepts, demonstrating a high reasoning ceiling.

The second, Tier 2 (Flawed but Occasionally Successful), includes {{mistral-nemo}} (8/20), and {{mistral}} (6/20). These models achieved several successes, proving they could handle basic translation and follow some instructions. However, this tier is critically flawed. Their successes, visible in Table \ref{tab:results_summary}, are overshadowed by major semantic failures where the code compiles but produces the wrong output, and a heavy reliance on C like syntax. These models often attempted to call {malloc()} or used C's {printf} format specifiers, revealing a shallow, syntactic understanding rather than deep semantic translation.

The final Tier 3 (Complete Failure) comprises the vast majority of models (14 of 19). Models such as {yi-coder}, {starcoder2}, {qwen3:14b}, {llama3.1:8b}, {granite-code:8b}, {gemma3:12b}, and {codellama} failed every single complex test, scoring 0. This comprehensive failure is documented in Table \ref{tab:results_summary}. Their primary failure, as detailed in Section \ref{sec:discussion}, is a fundamental inability to generate even the basic, runnable Java class structure required by the test harness, failing long before semantic translation could even be attempted.

\section{Discussion}
\label{sec:discussion}
The tiered performance in Section \ref{sec:results} is explained by a clear hierarchy of failure modes. Models in Tier 3 failed at the most basic level, while Tier 1 models failed only on the most complex semantic concepts. These failures can be categorized into four distinct groups.

The first and most common group was Fundamental Boilerplate \& Syntax Failure. This was the single most common reason for failure and plagued all 14 Tier 3 models, resulting in their 0 scores. These models were fundamentally incapable of generating a runnable Java file, failing at the first hurdle of the verification pipeline. A primary error was a {ClassNotFoundException}, as models like {gemma3}, {falcon3}, {granite-code}, {llama3.1}, {qwen3}, and {yi-coder} sometimes failed to generate a {public class} matching the filename. Even if a class was generated, many models failed to generate a {public static void main(String[] args)} method, making the class unrunnable (e.g., {mistral} on test 4, {llama3.1} on test 4). A common C like syntax was generating {public static int main(...)}, as seen from {llama3.1}, {gemma3}, and {opencoder}, which is invalid in Java and results in an unexpected return value error. Finally, many models produced non code hallucinations. {starcoder2} was the worst offender, outputting {**HINTS (IF NEEDED):**} directly into the code, while {dolphincoder}, {qwen3}, and {codellama} generated explanatory notes or pseudocode (e.g., {for line in lines:}), all causing immediate compilation failure.

Beyond this fundamental inability to generate valid files, the second category of failure was translating C syntax literally. These are failures where the model (typically Tier 2) produces runnable code, but directly copies C syntax or library calls that don't exist in Java. This leads to compilation or runtime errors and reveals a shallow, syntactic level understanding. A common example was boolean logic, where the C idiom {if (int\_variable)} was used instead of the Java compliant {if (int\_variable != 0)}, a mistake made by {qwen2.5-coder}, {phi4}, and {mistral}. Another frequent C syntax was related to memory management, with models like {codeqwen} and {deepseek-coder-v2} attempting to call C's {malloc()} or {free()} functions within the Java code. Models also failed by directly calling C library functions like {strcpy()} and {strlen()} ({codeqwen} on test 14) or by using C's {printf} format specifiers. The {\%u} (unsigned) specifier, for instance, was used by {phi4} and {mistral-nemo}, causing a runtime {UnknownFormatConversionException} in Java. Lastly, some models like {codellama} and {codeqwen} attempted to use C's address of operator ({\&}) or function pointer syntax directly in the Java output, resulting in compilation failures.

Perhaps the most dangerous failures were in the third category: Semantic \& Logic Failures (Output Mismatch). These are failures where the code compiles and runs but produces the wrong answer, indicating a deep logical flaw in the translation. These failures primarily affected the otherwise successful Tier 1 and Tier 2 models. For example, in test 2, {mistral-nemo} failed to replicate C's pass by value for structs, it passed the object reference directly, which was then modified by the callee, breaking the C code's logic of operating on a copy. In test 8, {deepseek-coder-v2} successfully translated the bitfield {struct} but failed to correctly implement the overflow behavior (truncation) for C bitfields, resulting in an output mismatch. Similarly, in test 13, {codeqwen} incorrectly translated a {goto} based loop, causing it to execute only once instead of five times. {Mistral} (test 20) also produced a semantic error by failing to correctly implement a {switch} statement with fall through, leading to a different final output than the C baseline.

Finally, the fourth group, Failures on Advanced C Concepts, demonstrates the hard ceiling of current capabilities, as even the best Tier 1 models struggled with these. {Enum} translation (test 12) was universally failed, models either didn't create an {enum} at all (leading to {cannot find symbol: STOPPED}) or defined it inside the {main} method, causing an {enum types must not be local} error. Global variables (test 16) were frequently and incorrectly translated as {public static final}, making them unmodifiable and failing the test which required mutation. Function pointers (test 7) were a near total failure (0\% success), as most models did not know to use Java's functional interfaces (like {IntBinaryOperator}) and failed to add the required {import java.util.function.\*;}. Lastly, {sizeof} (test 18) was handled incorrectly by every model, {phi4} produced the wrong size, while others hallucinated a {sizeof()} function or failed compilation ({cannot find symbol: Pointer.SIZE}).

\section{Conclusion}
In this paper, we presented REMODEL-LLM, a hybrid AST driven pipeline, to empirically assess the capability of 19 small, quantized Edge LLMs on the complex C to Java translation task. Our findings, based on the comprehensive results in Table \ref{tab:results_summary}, reveal a stark, multi tiered hierarchy of model competence. The vast majority (14 of 19 models) are in Tier 3, demonstrating a 0\% success rate. These models failed at the most basic level, unable to generate the fundamental Java class boilerplate, and were plagued by errors like {ClassNotFoundException}, missing {main} methods, and code block hallucinations. A small Tier 2 ({mistral-nemo}, {mistral}) was able to produce runnable code but was critically flawed, producing dangerous semantic errors (e.g., incorrect {goto} loop logic) and naively translating C specific syntax like {printf} format specifiers and {malloc} calls.

Only three Tier 1 models, {codeqwen}, phi4 and deepseek-coder-v2, proved viable, passing over 50\% of the tests. These models demonstrated a unique ability to follow the complex, rule based instructions in our prompts. However, even these top tier models hit a hard ceiling. As shown in Table \ref{tab:test_case_analysis}, all models, including the best, failed on the most semantically complex C concepts: {enum} logic, {sizeof} simulation, and {int**} refactoring. Most notably, the 0\% pass rate on function pointers shows a universal inability to bridge the C to Java paradigm gap by introducing Java's functional interfaces. This failure spectrum from basic syntax to advanced logic provides a clear map of the reasoning capabilities of current models. It shows that while a select few are approaching viability, the problem of fully automated, semantically correct C to Java translation remains unsolved.

For future work, several promising avenues exist to build upon this research. First, fine tuning a Tier 1 model like {deepseek-coder-v2} on a large, high quality C to Java parallel corpus could bake in the complex rules that our prompts (Listings \ref{lst:func_prompt}, \ref{lst:global_prompt}, \ref{lst:struct_prompt}) attempt to provide at inference time, potentially improving both accuracy and consistency. Second, an iterative self correction pipeline could be developed, this system would feed the {javac} compilation error or the runtime {stdout} mismatch back to the LLM in a second pass prompt, asking it to fix its own mistake, thereby creating a reasoning loop. Finally, a multi pass refactoring approach could be explored, where a first pass performs a naive, syntactic translation and a second, specialized refactoring pass, armed with the full class context, attempts to solve the Impossible challenges like function pointers and {int**} parameters. 

We must also acknowledge a significant threat to external validity, which places a clear boundary on our conclusions. This study's evaluation is built upon a 20 case, single file micro benchmark where each test is intentionally designed to isolate a specific C idiom. This approach is necessary for pinpointing the exact semantic reasoning failures of each model, but it does not reflect the complexity of production scale legacy systems. Real world C codebases are typically multi file repositories managed by complex build systems (e.g., make), which rely heavily on header files, macros, and external definitions to manage a global program state. The true challenge of the task lies not just in translating these idioms in isolation, but in handling their combinatorial explosion, for example a function pointer (T7) nested within a struct (T19) that is passed by value (T2). Our current REMODEL-LLM pipeline, which operates on a file by file basis, is not yet equipped to parse or resolve this network of cross file dependencies. Therefore, while our methodology is valid for establishing a foundational baseline of what is possible with current quantized models on discrete reasoning tasks, its direct, real world applicability is limited. We have established a clear performance ceiling on these isolated problems, but scaling this approach to a full, interdependent C project remains a significant and open challenge.

\section*{Data and Code Availability}
The complete dataset, including all C test cases, the Python pipeline code (CAnalyzer, JavaTranslator, etc.), and the raw output logs for all 19 models are available at the following GitHub repository: \href{https://github.com/arygup/REMODEL-LLM-Refactoring-and-Modernizing-Existing-Legacy-Code-using-Quantized-LLMs}{https://github.com/arygup/REMODEL-LLM-Refactoring-and-Modernizing-Existing-Legacy-Code-using-Quantized-LLMs}

\section*{Acknowledgment}
The authors would like to thank the developers of the {ollama} framework and the research teams behind the open-source models that made this study possible.

\bibliography{references}

\end{document}